\documentclass[pre,aps,floats,twocolumn,superscriptaddress,floatfix,article]{revtex4-1}
\usepackage{graphicx,amssymb,amsmath,ifthen}

\usepackage{natbib}

\usepackage{times}
\usepackage{bm}

\begin{document}
\title{Majorana bound states in non-homogeneous semiconductor nanowires}

\author{Christopher Moore}
\affiliation{Department of Physics and Astronomy, Clemson University, Clemson, SC 29634, USA}

\author{Tudor D. Stanescu}
\affiliation{Department of Physics and Astronomy, West Virginia University, Morgantown, WV 26506, USA}

\author{Sumanta Tewari}
\affiliation{Department of Physics and Astronomy, Clemson University, Clemson, SC 29634, USA}

\begin{abstract}
We demonstrate that partially overlapping Majorana bound states (MBSs) represent a generic low-energy feature that  emerges in non-homogeneous semiconductor nanowires coupled to superconductors in the presence of a Zeeman field. The emergence of these low-energy modes is not correlated with any topological quantum phase transition that the system may undergo as the Zeeman field and other control parameters are varied. Increasing the characteristic length scale of the variations in the potential leads to a continuous evolution from strongly overlapping MBSs, which can be viewed as ``regular'' Andreev bound states (ABSs) that cross zero energy, to well separated weakly overlapping MBSs, which have nearly zero energy in a significant range of parameters and generate signatures similar to the non-degenerate zero-energy Majorana zero modes (MZMs) that emerge in the topological superconducting phase. We show that using charge (or spin) transport measurements it is virtually impossible to distinguish MZMs from weakly overlapping MBSs emerging in the topologically-trivial regime.
\end{abstract}
\maketitle

Topological superconductors, like topological insulators, are characterized by a gap in the bulk excitation spectrum to fermionic excitations, but gapless excitations on the boundary \cite{Read_Green_2000,Kitaev_2001,Nayak_2008}. In one or quasi-one-dimensional systems the boundaries are the two edges and the gapless excitations are localized near the end points. In topological superconductors, the second quantized operators corresponding to the edge-localized, non-degenerate, zero-energy quasi-particles satisfy the property $\gamma^{\dagger}=\gamma$. This property, which identifies particles with their own anti-particles, was first proposed by E. Majorana in 1937 \cite{Majorana} in the context of high energy physics as allowable solution to the Dirac equation. In topological superconductors the non degenerate zero energy edge modes, also called Majorana zero modes (MZMs), are Bogoliubov excitations, and can be viewed as an equal amplitude admixture of particle- and hole-like components of Bogoliubov-de Gennes (BdG) wave functions.

While MZMs have not yet been conclusively observed in experiments, they have been theoretically shown to exist in low dimensional spinless $p$-wave superconductors \cite{Read_Green_2000,Kitaev_2001}, as well as in similar systems such as topological insulators with proximity induced superconductivity \cite{Fu_2008}, and spin-orbit-coupled semiconductor-superconductor heterostructures \cite{Sau,Annals,Alicea,Long-PRB,Roman,Oreg,Stanescu}.
In particular, the semiconductor-superconductor heterostructure, involving a low-dimensional semiconductor with spin-orbit coupling in proximity to a $s$-wave superconductor and an externally applied Zeeman field, and a system of chains of magnetic adatoms on the surface of a bulk superconductor with spin-orbit coupling, have motivated tremendous experimental efforts with a number of recent works claiming to have observed signatures consistent with MZMs \cite{Mourik_2012,Deng_2012,Das_2012,Rokhinson_2012,Churchill_2013,Finck_2013,Perge,Ruby}. More recently, superconducting heterostrucures fabricated with a semiconducting core and an epitaxial superconducting shell have been shown to exhibit a high quality proximity effect allowing researchers to measure charging effects in the Coulomb blockade regime \cite{Marcus}. Zero bias conductance peaks in these nanowires with normal leads attached at each end in the Coulomb blockade regime (the so-called teleportation signal \cite{Fu}), and exponential suppression of energy splitting in the ground state energy modes with increasing wire length have been cited as strong evidence of the existence of MBSs \cite{Marcus,Heck}.

\begin{figure}
	\begin{center}
		\includegraphics[width=75mm]{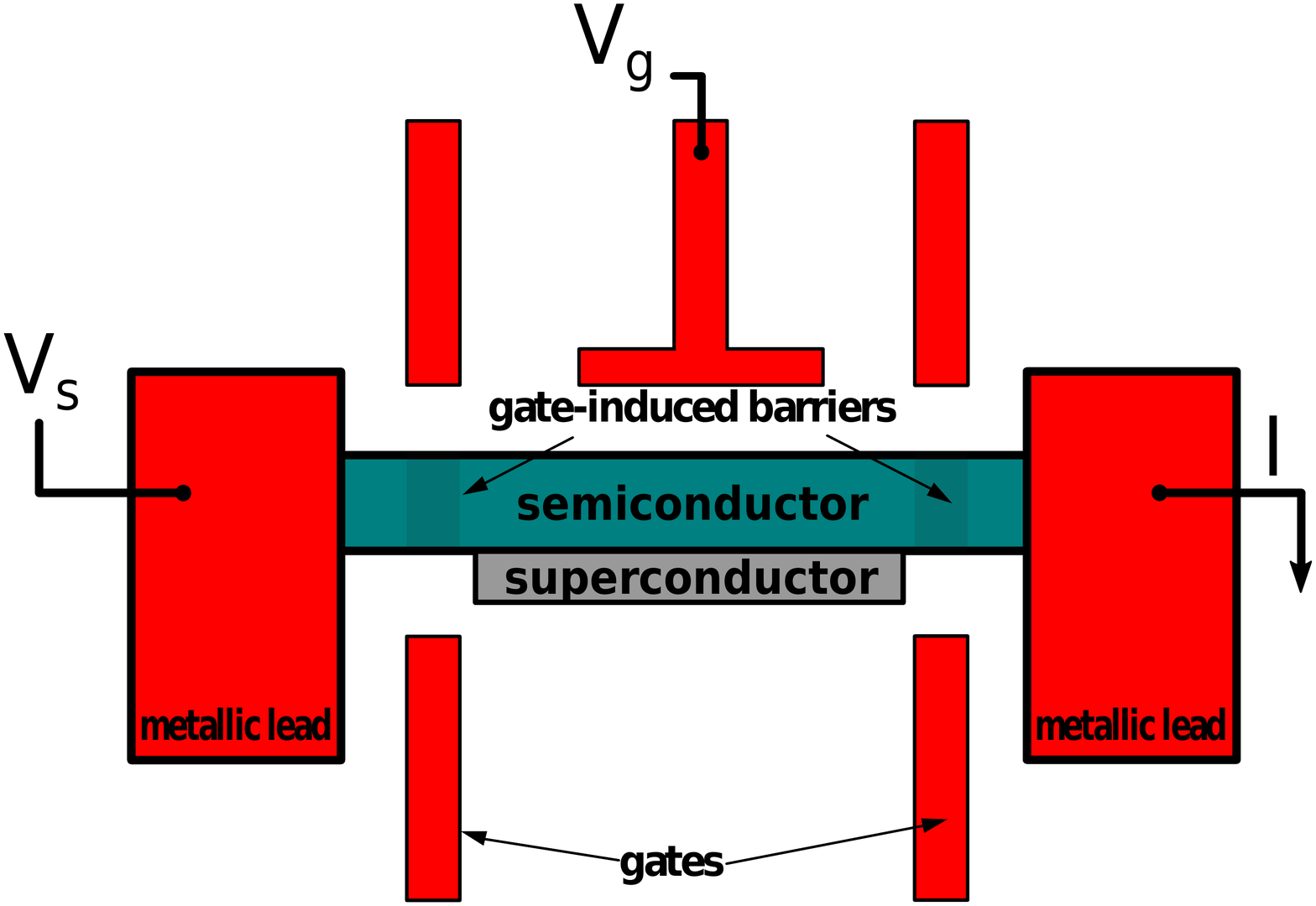}
	\end{center}
	\caption{Diagram of the experimental setup.  Metallic leads are coupled to each end of a semiconductor nanowire proximity-coupled to an s-wave superconductor. Potential gates create tunneling barriers and control the electrostatic energy of the heterostructure. A bias voltage $V_s$ is applied across the wire generating a current $I$.} 
	\label{fig:smothConfine}
\end{figure}

\begin{figure}
	\begin{flushright}	
		\includegraphics[width=72mm]{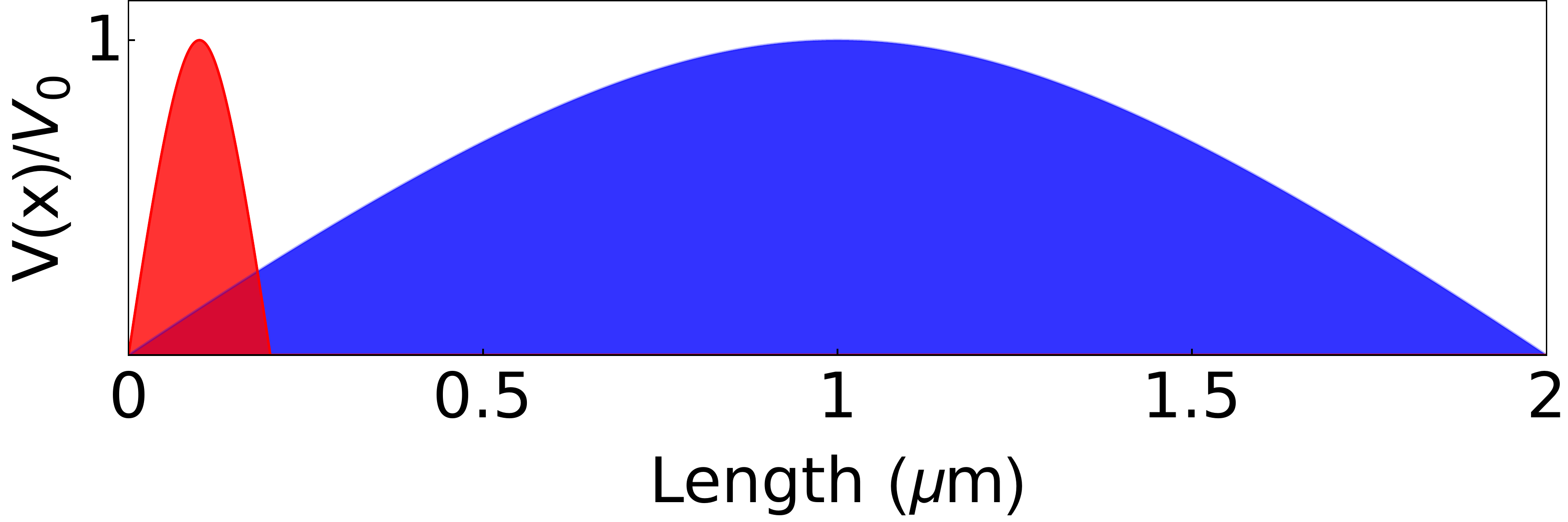}\llap{
			\parbox[b]{165mm}{\large\textbf{(A)}\\\rule{0ex}{20mm}}}	
		\includegraphics[width=75mm]{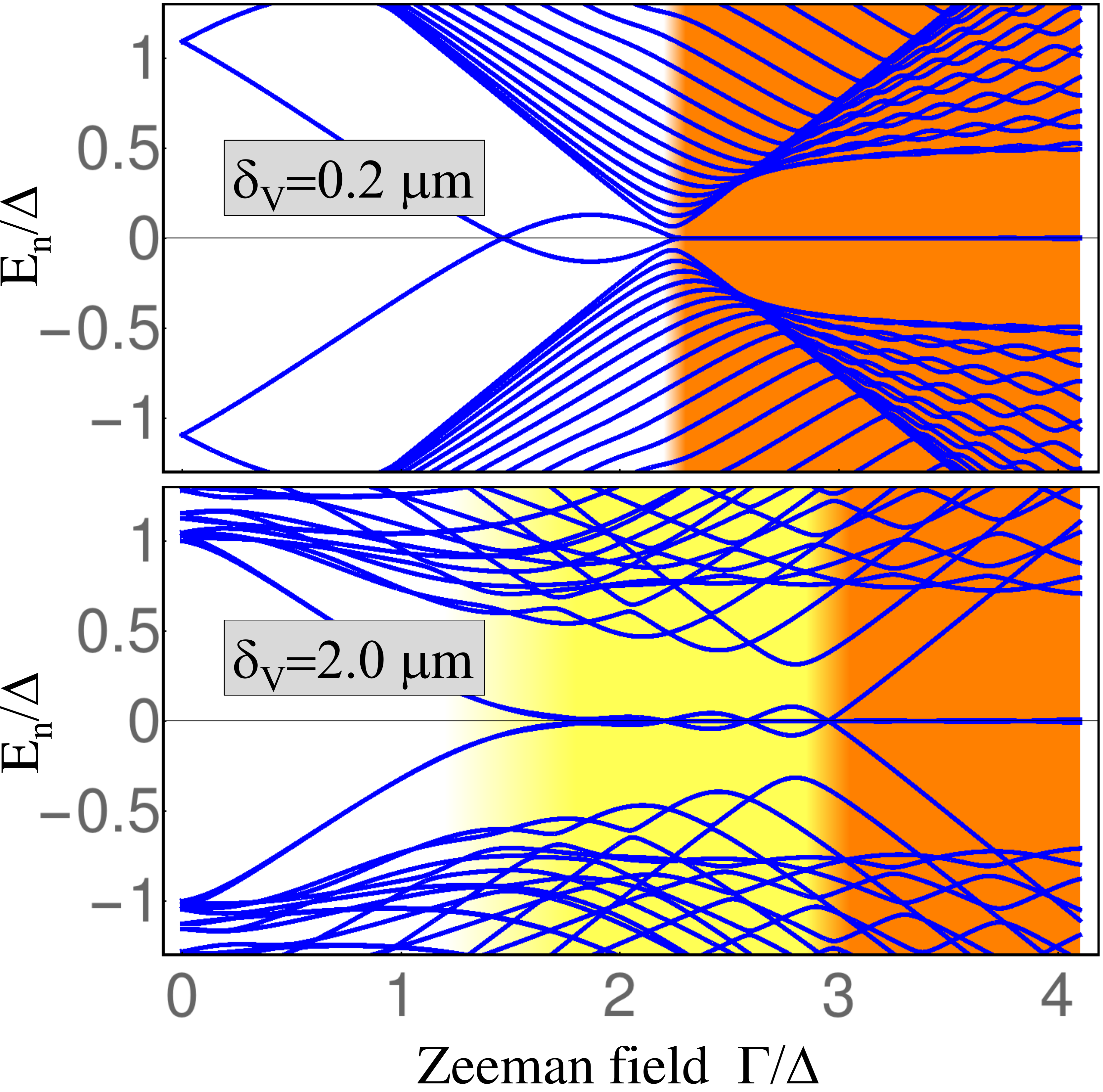}\llap{
			\parbox[b]{165mm}{\large\textbf{(B)}\\\rule{0ex}{30mm}
				\large\textbf{(C)}\\\rule{0ex}{37mm}}}				
	\end{flushright}
	\caption{(A) Non-homogeneous potential profiles with characteristic widths $\delta_V = 0.2~ \mu$m (red) and $\delta_V =2.0~\mu$m (blue). The maximum height/depth of the potential  is $V_0$, which can be positive or negative. (B) and (C)  Zeeman field dependence of low-energy spectra.   (B)  A short-range potential inhomogeneity induces an ABS that crosses zero energy in the topologically-trivial region. A robust MZM emerges in the topological regime (orange). (C) A long-range inhomogeneity generates four nearly-zero energy ABSs in the topologically-trivial regime (yellow).}
	\label{fig:FFig4}
\end{figure}

\begin{figure}
	\begin{flushright}
		\includegraphics[width=80mm]{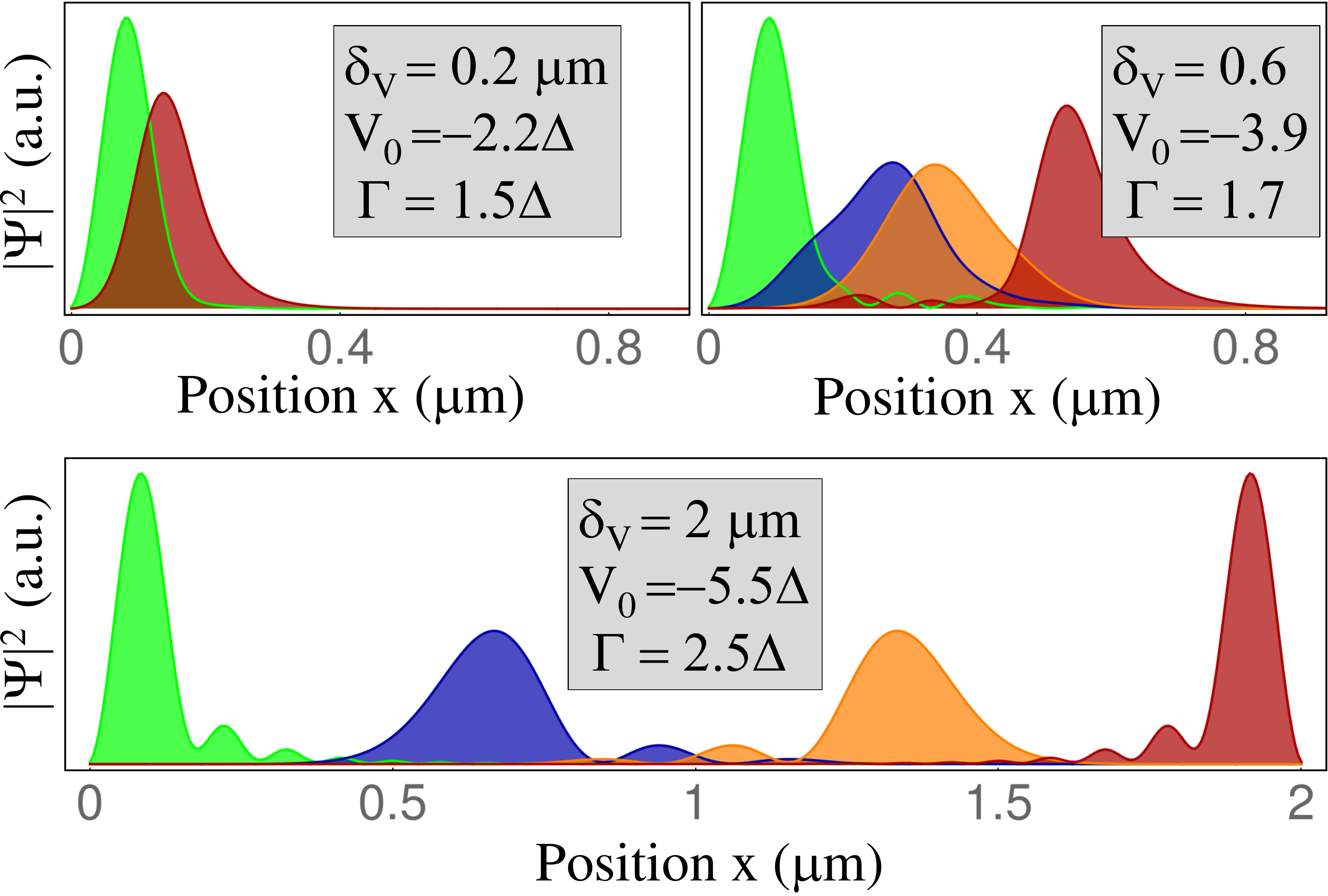}\llap{
			\parbox[b]{130mm}{\large\textbf{(A)}\\\rule{0ex}{27mm}
				\large\textbf{(C)}\\\rule{0ex}{21mm}}}\llap{
			\parbox[b]{55mm}{\large\textbf{(B)}\\\rule{0ex}{49.2mm}
			}}
			
		\end{flushright}
		\caption{Majorana wave functions for the  lowest energy modes in the topologically trivial regime. (A) A localized potential inhomogeneity generates two strongly overlapping MBSs, which correspond to a regular ABS. (B) Intermediate-range inhomogeneity. The lowest energy MBSs become well separated and another pair of MBSs starts to separate spatially. (C) Four weakly overlapping MBSs in a system with long-range potential inhomogeneity.}
		\label{fig:wfPot}
	\end{figure}

In this work we first show that, in non-homogeneous semiconductors, short length scale potential inhomogeneities can generically give rise to zero energy localized Andreev bound states, independent of whether the system is topological or trivial. This result contradicts an earlier result for purely local potentials \cite{Sau_Demler}, where such zero energy resonance was found only for the parameters appropriate for the topological phase. We find that, while for short length scale inhomogeneity the zero energy states are unstable to changes in Zeeman fields and amplitude of the potential inhomogeneity, for potential variations of longer length scales they cross over to sub-gap zero energy resonance that are surprisingly robust to perturbations (see Fig. \ref{fig:FFig4}), \textit{even if the system is topologically trivial}. The crossover takes place via a gradual unfolding of the Andreev bound states, which can be viewed as a pair of strongly overlapping Majorana bound states (MBSs), into a pair of spatially separated weakly overlapping MBSs still in the effectively topologically trivial phase, with increasing length scale of potential variations (see Fig. \ref{fig:wfPot}).

This result is important because the charge densities in the experimental semiconductor-superconductor heterostructures are controlled by multiple suitably placed gate potentials, which can generically cause potential variations. The short or long range potential inhomogeneities and partially unfolded robust Andreev states are thus expected to be quite generic in experiments. By calculating the local density of states (LDOS), zero bias conductance peaks, and teleportation amplitude in the Coulomb blockade regime we show that, in such wires, it is virtually impossible to distinguish the mutually overlapping pairs of MBSs at each end in the topologically trivial phase from spatially well separated non-degenerate MZMs localized near the ends, the hallmark of topological superconductors, in any local measurements, in the absence of an interferometric signal \cite{Sau_Swingle_Tewari}.

The low-energy physics of semiconductor-superconductor heterostructures is investigated using a Bogoliubov de Gennes (BdG) formalism based on an effective  tight binding  Hamiltonian of the form
\begin{subequations}\label{eq:7}
\begin{align}
	H =&  -t_x\sum_{i,\delta_x,\sigma}c_{i+\delta_x\sigma}^\dagger
	c_{i\sigma}-t_y\sum_{i,\delta_y,\sigma}c_{i+\delta_y\sigma}^\dagger
	c_{i\sigma}\\
	&-\mu\sum_{i,\sigma}c_{i\sigma}^{\dagger}c_{i\sigma}+ \Gamma\sum_{i}c_{i}^{\dagger}\sigma_x c_i \\
	& +\frac{i}{2}\sum_{i,\delta}\left[\alpha c_{i+\delta_x}^{\dagger}\sigma_y c_i-\alpha_y c_{i+\delta_y}^{\dagger}\sigma_x c_i +h.c.\right]\\
	&+\sum_{i}\Delta_i\left(c_{i\uparrow}^{\dagger}c_{i\downarrow}^{\dagger}+h.c.\right)\tau_x+\sum_{i,\sigma}V_c\left(i\right)c_{i\sigma}^{\dagger}c_{i\sigma},
\end{align}
\end{subequations}
where the lattice sites $i$ correspond to $N_y$ parallel chains oriented along the x-direction,  $t_x$ and $t_y$ are hopping matrix elements, $\mu$ is the chemical potential, $\Gamma$ the Zeeman field, $\alpha$ and $\alpha_y$ the longitudinal and transverse Rashba coeficients, respectively, and $\Delta_i$ is the induced pair potential. The non-homogeneous background potential is described by the position-dependent function $V_c(i)$. Typical potential profiles used in the  calculation are shown in Fig. \ref{fig:FFig4} (A).


The low-energy spectrum is obtained by numerically diagonalizing the BdG Hamiltonian. In addition, we calculate the tunneling differential conductance in the single-lead and two-lead configurations. Note that in the single-lead configuration the current $I$ is extracted through the superconductor, while in the two-lead setup the superconductor is either grounded or isolated, the last case corresponding to Fig. \ref{fig:smothConfine}.  The zero temperature differential conductance,  $G_0(\varepsilon)=dI/dV$,  is calculated using the Landauer formula
\begin{equation}\label{eq:conductance}
G_0(\varepsilon)=\frac{e^2}{\pi\hbar}\sum_{n}T_n(\varepsilon),
\end{equation}
where $\varepsilon$ is the onsite energy in the wire and $T_n$ are the transmission eigenvalues for each conduction channel.  
The conductance at finite temperature is
\begin{equation}\label{eq:19}
G\left(V,T\right)=\int d\varepsilon G_0\left(\varepsilon\right)\frac{1}{4T\cosh^2\left[\left(V-\varepsilon\right)/2T\right]},
\end{equation}
where V is the bias voltage. A temperature $T\approx 100$ mK was used in the numerical calculations. In addition, the values of the hopping parameters were chosen to correspond to an effective mass $m^*=0.04m_0$, the Rashba coefficient is $\alpha = 200$ meV$\cdot$\AA, the induced gap is  $\Delta=0.25$ meV, and the wire length $L=2~\mu$m.

The emergence of potential-induced low-energy modes in the topologically-trivial regime is illustrated in Fig. \ref{fig:FFig4}. We emphasize that, when discussing finite systems, the trivial and topological ``phases'' have to be defined operationally. For example, we will call ``topological phase'' the regime characterized by the presence of two  MZMs (associated with the same spin-split sub-band) localized at the ends of the wire. Also, the topological ``phase transition'' is, in fact, a crossover signaled by a minimum (rather than a zero) of the bulk gap.
First, we consider a short-range potential corresponding to the red curve in Fig.  \ref{fig:FFig4}(A). The potential induces an Andreev bound state that crosses zero energy in the topologically trivial regime, as shown in panel (B). Note that this is in contrast with the behavior generated by a purely local potential (i.e., with $\delta_V=0$), which was shown \cite{Sau_Demler} to induce low-energy modes {\em only} in the {\em topological} regime. Next, we consider the potential corresponding to the blue curve in Fig.  \ref{fig:FFig4}(A). In this case, upon increasing the Zeeman field four low-energy modes emerge while the system is still in the topologically-trivial regime, as shown in panel (C).

The partially-overlapping MBSs responsible for the low-energy modes discussed above are shown in Fig. \ref{fig:wfPot}. The ABS induced by the localized potential (see Fig. \ref{fig:FFig4}B) corresponds to two strongly overlapping MBSs, as shown in panel \ref{fig:wfPot}A. Increasing the width of the potential results in a larger spatial separation between the two modes. In addition, another pair of low-energy overlapping MBSs  emerges, as illustrated in panel \ref{fig:wfPot}B. Finally,  the low-energy bound states populating the yellow (light gray) trivial regime in Fig. \ref{fig:FFig4}C are the four weakly overlapping Majorana modes shown in panel \ref{fig:wfPot}C. In general, if the effective potential in the semiconductor wire has variations of the order of the induced gap on length scales comparable to the Majorana localization length, the low-energy modes correspond to a ``Majorana chain'' similar to the situation illustrated in Fig. \ref{fig:wfPot}C. Note that these MBSs can be associated with either one of the top occupied spin-split sub-bands.

\begin{figure}
	\includegraphics[width=75mm]{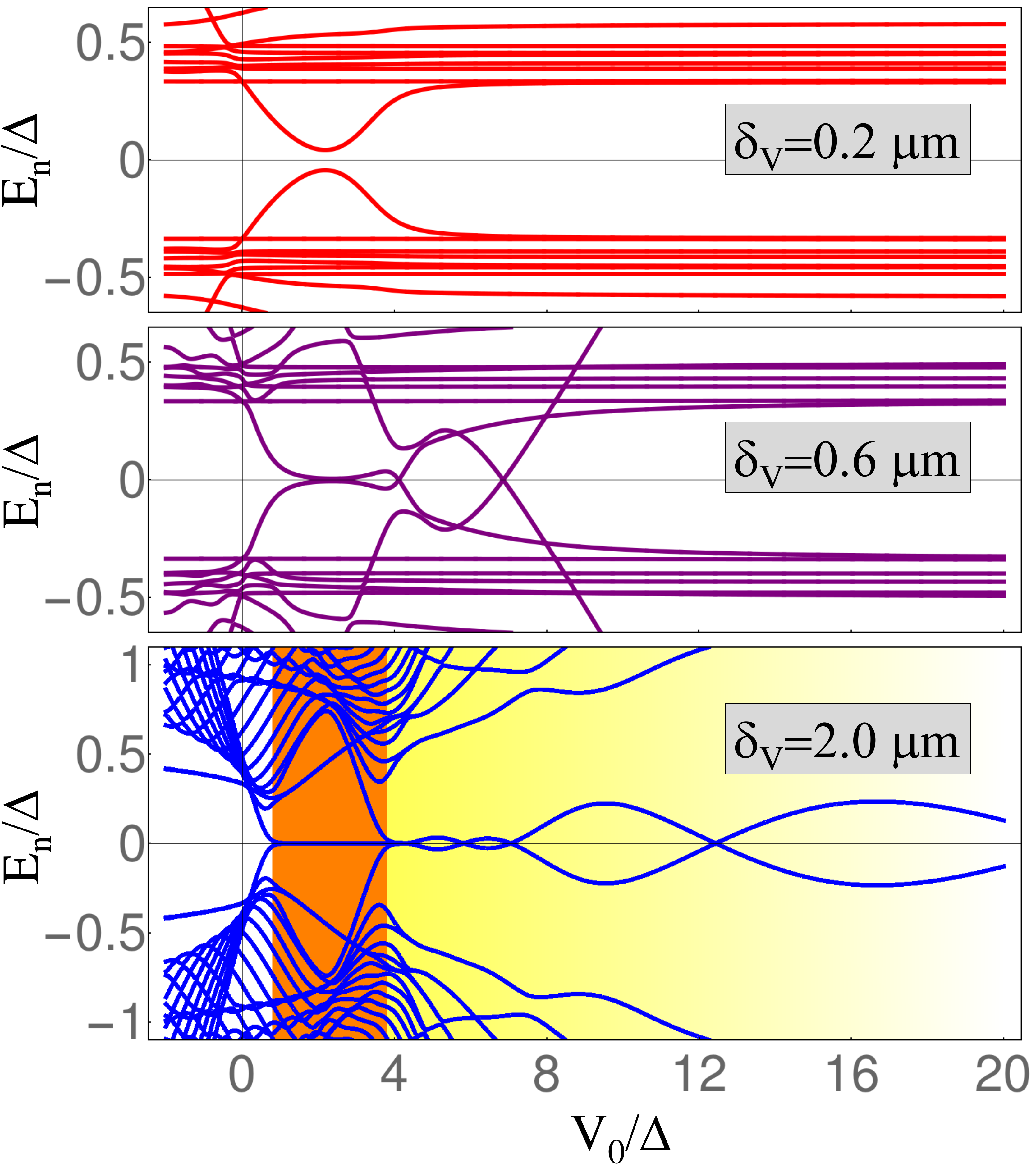}\llap{
		\parbox[b]{155mm}{\large\textbf{(A)}\\\rule{0ex}{22mm}
			\large\textbf{(B)}\\\rule{0ex}{22mm}	\large\textbf{(C)}\\\rule{0ex}{32mm}}}
	\includegraphics[width=75mm]{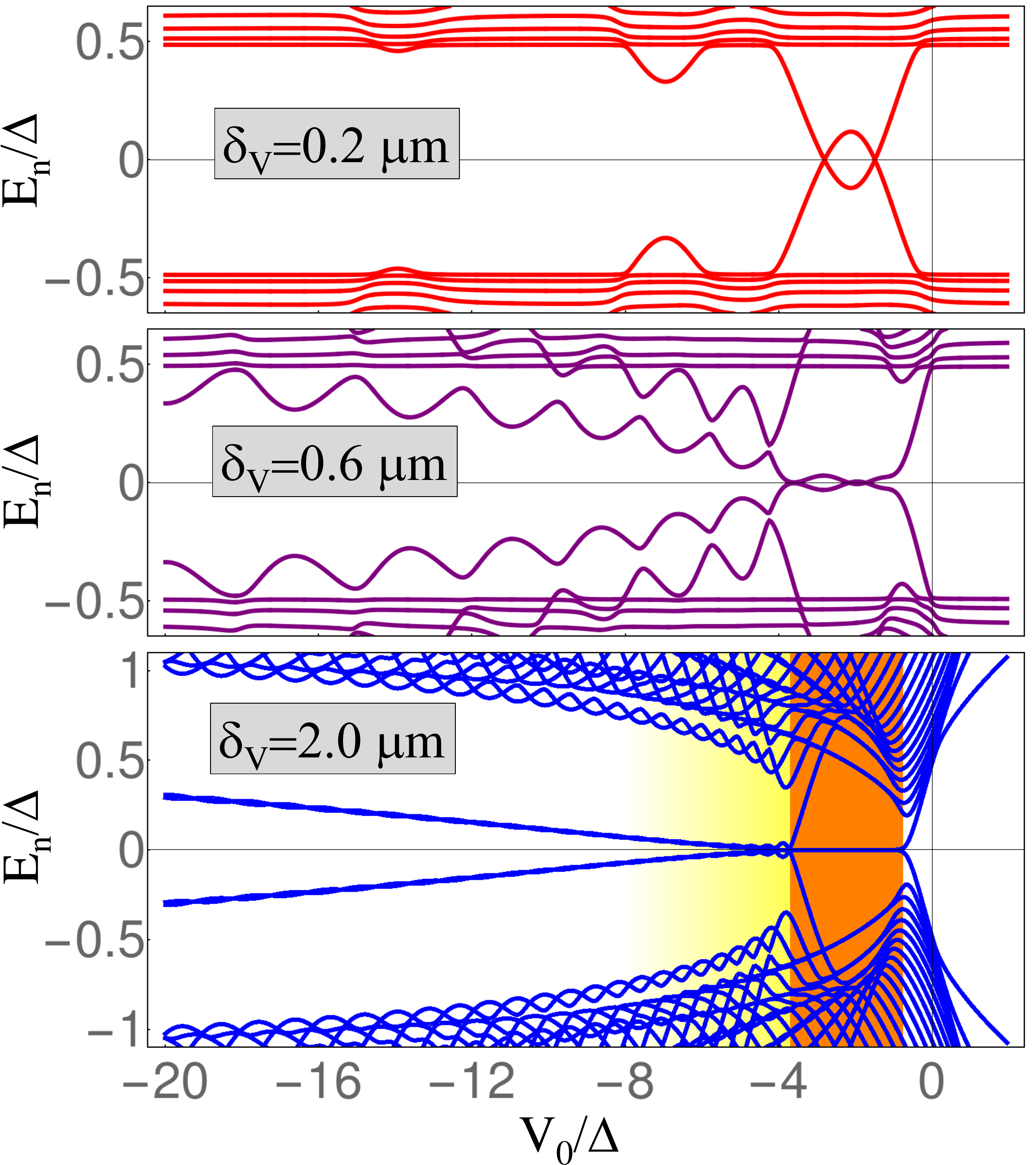}\llap{
		\parbox[b]{155mm}{\large\textbf{(D)}\\\rule{0ex}{22mm}
			\large\textbf{(E)}\\\rule{0ex}{22mm}	\large\textbf{(F)}\\\rule{0ex}{32mm}}}	
	\caption{Low-energy spectra in the presence of a non-homogeneous potential. The Zeeman field is $\Gamma=1.75\Delta$ and the chemical potential (defined relative to the bottom of the top band when $V_0=0$) is $\mu=2\Delta$ in panels (A-C) and $\mu=-2\Delta$ in panels (D-F). A finite-range localized potential ($\delta_V>0$) always generates a zero-crossing mode for $V_0<0$ [see panel (D)]. As the characteristic length scale $\delta_V$ increases, robust low-energy modes emerge in the topologically-trivial regime [yellow (light gray) regions in panels (C) and (F)].}
	\label{fig:FFig2}	
\end{figure}

The dependence of the low-energy modes associated with the partially overlapping MBSs on the inhomogeneous  potential is shown in Fig. \ref{fig:FFig2}. For localized potentials, $\delta_V\ll L$, the parameters of the calculation ($\Gamma=1.75\Delta$ and $|\mu|=2\Delta$) correspond to the topologically-trivial regime. First, we note that for any finite-range localized potential ($\delta_V>0$) a low-energy mode that crosses zero energy, which is associated with a pair of strongly overlapping MBSs, always emerges in the trivial regime for some $V_0<0$. Increasing the characteristic length scale $\delta_V$ of the potential inhomogeneity stabilizes these low-energy modes [see panels (B) and (E)] as a result of increasing the spatial separation between the MBSs. For large-enough values of $\delta_V$, {\em effectively zero-energy} Majorana modes emerge in both the topological (orange/dark gray) and trivial (yellow/light gray) regimes [see panels (C) and (F)].



\begin{figure}
	\begin{flushright}
		\includegraphics[width=80mm,height=33mm]{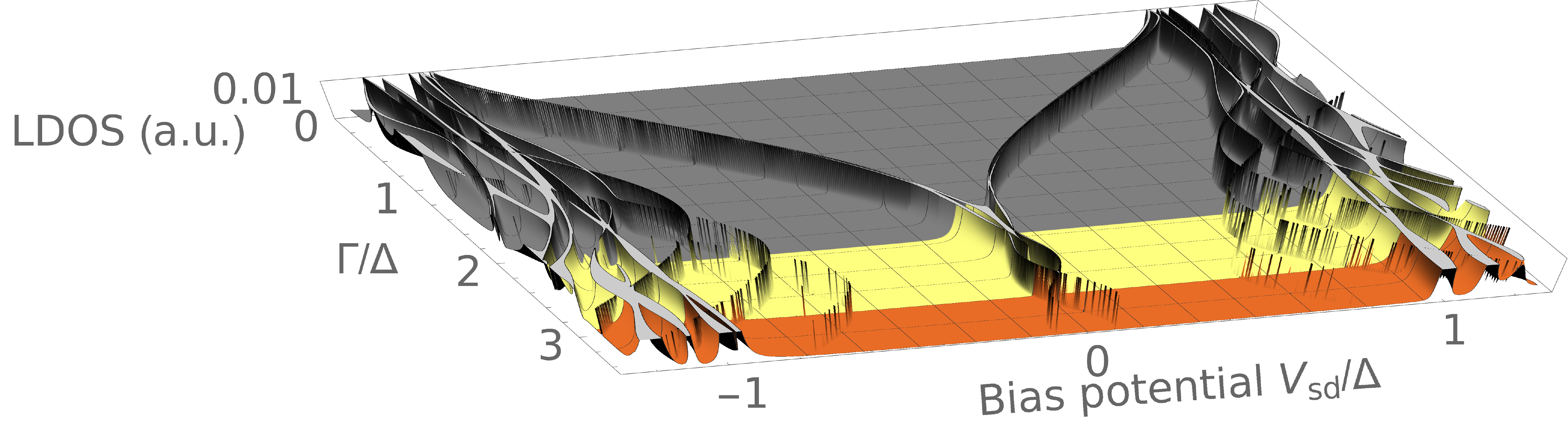}
		\includegraphics[width=75mm]{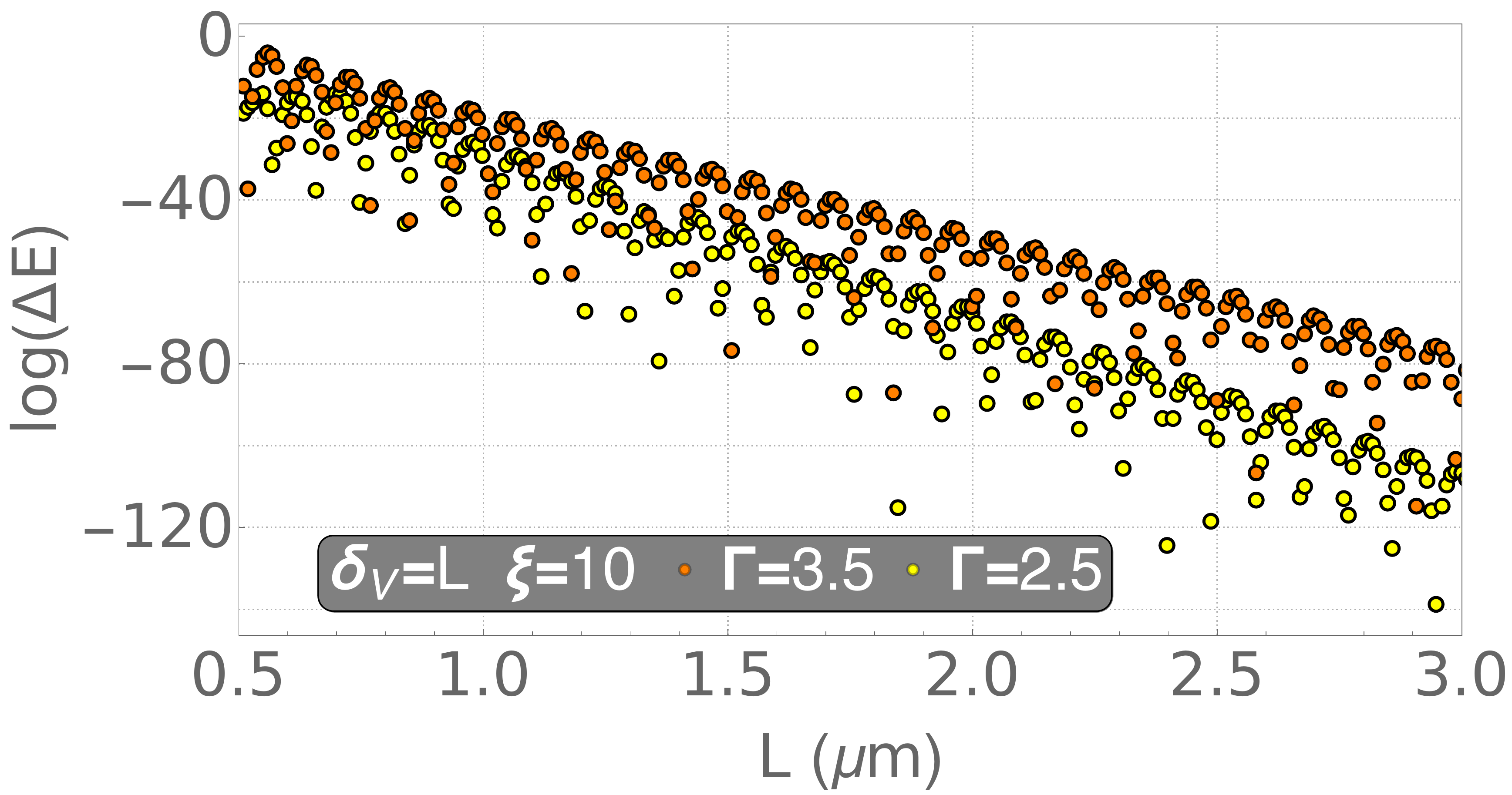}				
	\end{flushright}
	\caption{(top) Measurements of LDOS as a function of bias potential and Zeeman splitting associated with a potential length of $\delta_{V}=2.0\mu m$. For sufficiently long potential length scales zero energy crossings in the LDOS spectrum exist in the topologically trivial regime (yellow). (bottom) Logarithm of the energy splitting $\Delta E$ as a function of wire length $L$ showing exponential protection of the energy splitting. Energy splitting $\Delta E\propto\exp(-L/\xi)$ is numerically calculated in both the topologically trivial regime (yellow dots, $\Gamma=2.5$) and the topological regime (orange dots, $\Gamma=3.5$) with a Majorana decay length $\xi=10$ and potential length scale $\delta_V$ equal to that of the length of the wire. (periodic "beading" in the exponential decay is due to constructive and destructive interference between the MBSs \cite{Dumitrescu}.)}
	\label{fig:LDOS}	
\end{figure}

\begin{figure}
	\begin{center}
		\includegraphics[width=65mm]{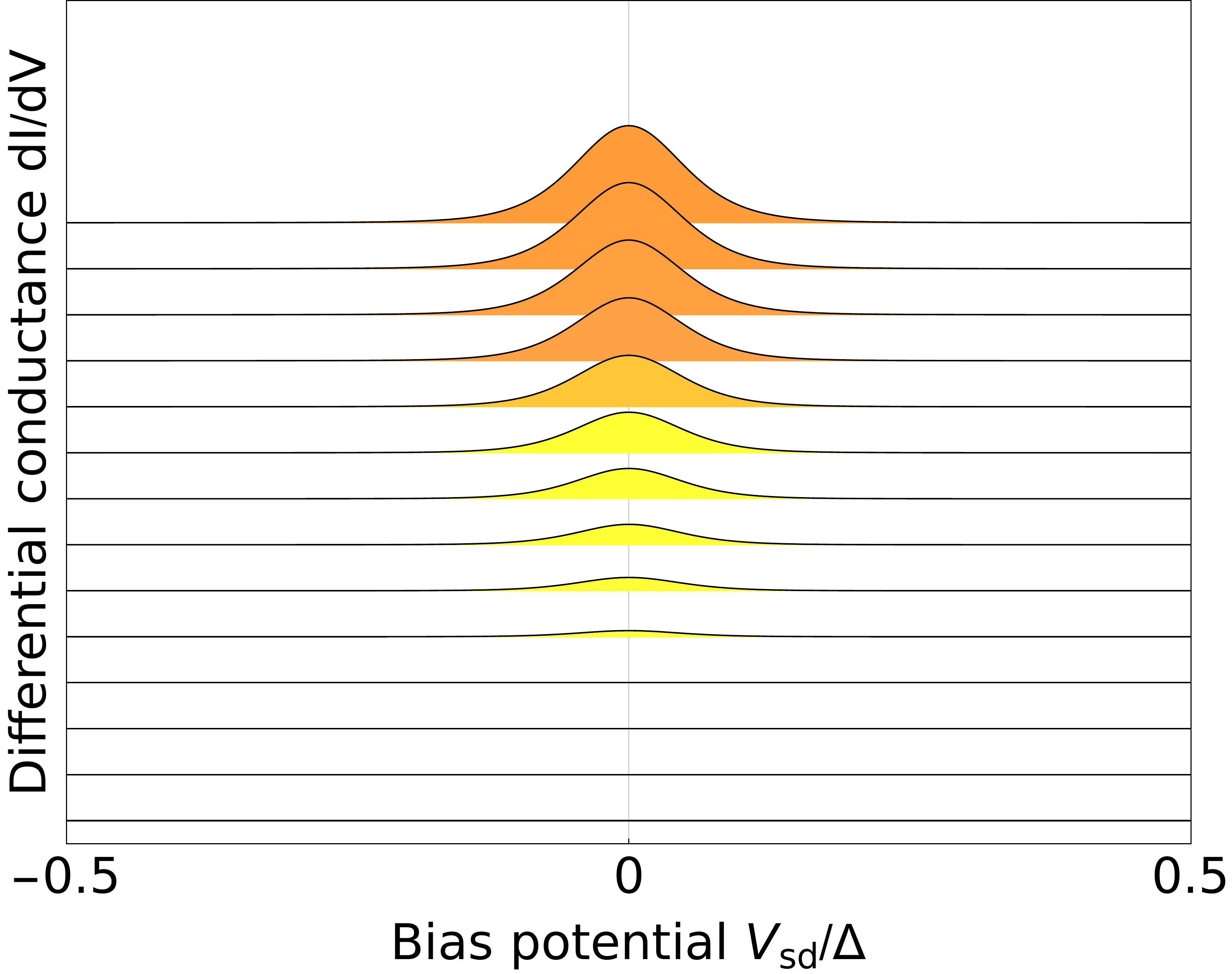}\llap{
			\parbox[b]{145mm}{\large\textbf{(A)}\\\rule{0ex}{45mm}}}
		\includegraphics[width=65mm]{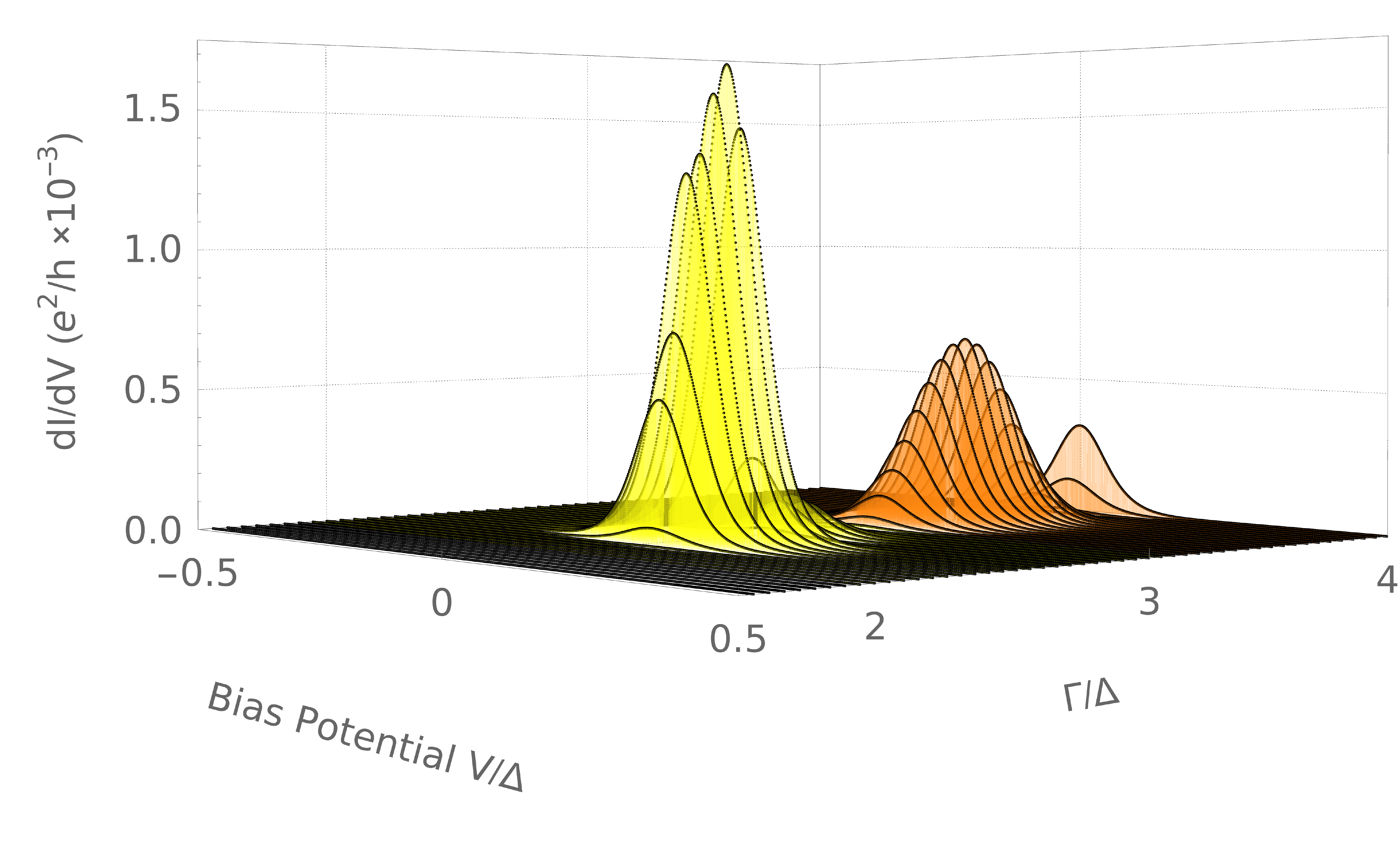}\llap{
			\parbox[b]{145mm}{\large\textbf{(B)}\\\rule{0ex}{30mm}}}
		\includegraphics[width=65mm]{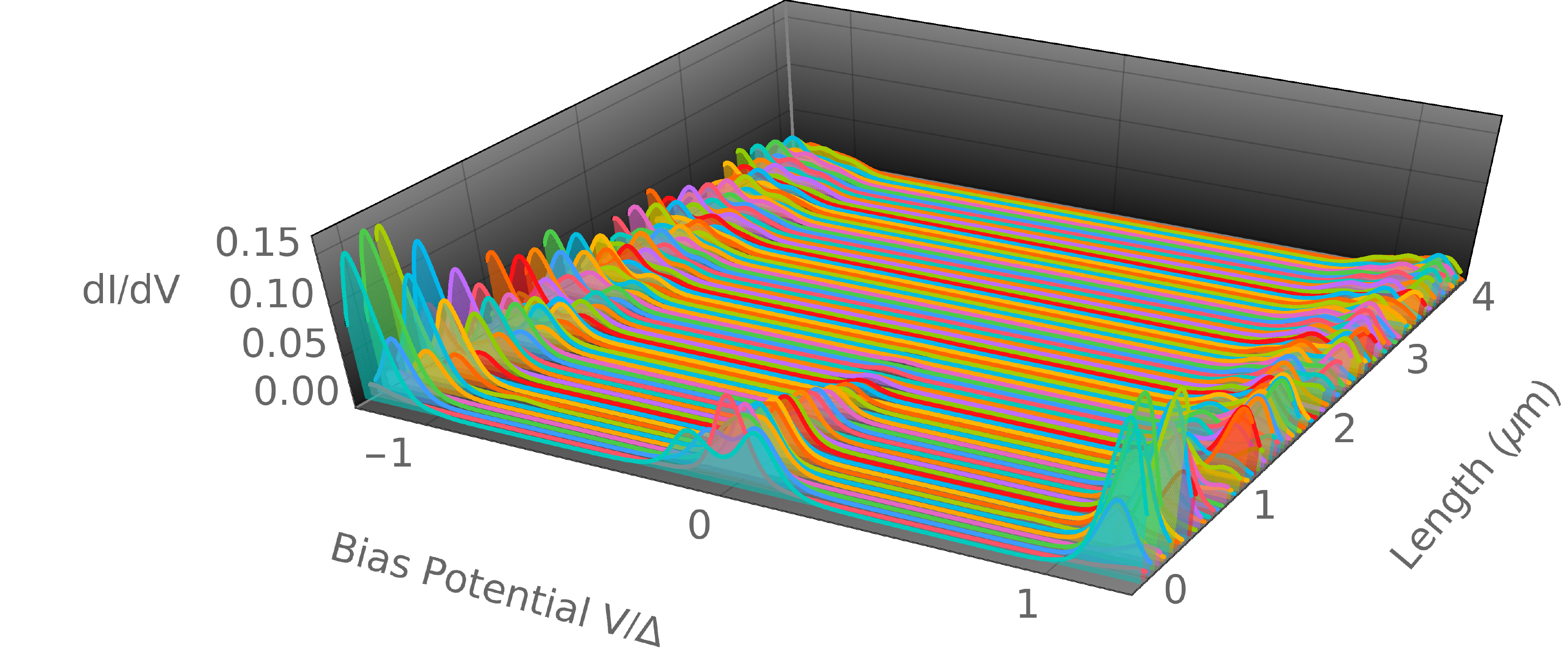}\llap{
			\parbox[b]{145mm}{\large\textbf{(C)}\\\rule{0ex}{25mm}}}	
	\end{center}
	\caption{(A) Single lead differential conductance plot of a $2\mu m$ superconductor-semiconductor heterostructure with a normal lead at one end of the wire. The Zeeman field ranges from $\Gamma=1.5\Delta$ to $\Gamma=4\Delta$ with profiles offset for clarity. (B) Two lead differential conductance in the Coulomb blockade regime  as a function of Bias potential $V$ and Zeeman splitting $\Gamma$ with long length scale potential fluctuations $\delta_V=2.0\mu m$ . Peaks exist in both the topologically trivial (yellow)  and the topological (orange) regimes.  These peaks are suppressed at 0 energy crossings in the spectrum. (C) Two lead teleportation differential conductance as a function of bias potential and length of wire at $\Gamma = 4\Delta$. As the length of the wire increases the height of the conductance peak rapidly decreases to zero.}
	\label{fig:cond}
\end{figure}

Given the ubiquity of the low-energy modes associated with partially overlapping MBSs in non-homogeneous systems, the following key question arises: how can one distinguish experimentally between partially overlapping MBSs, which is a robust near zero energy excitation descending from topologically trivial Andreev bound states, and spatially well separated, non-degenerate, zero-energy MZMs associated with the topological phase, where ``trivial'' and ``topological'' have well-specified operational definitions described earlier? We find that partially overlapping robust Majorana bound states located in the trivial phase are indistinguishable from topological MZMs localized near the ends using localized conductance measurements.  As shown in Fig. \ref{fig:LDOS}, the local density of states (LDOS) has a pronounced peak at zero energy at both the topological regime (orange) with single MZM localized near each end, as well as  in the effectively non-topological regime (yellow) where a pair of partially overlapping robust MBSs exist near each end (Fig. \ref{fig:wfPot}C) as a result of long length scale potential inhomogeneity. In Fig.~\ref{fig:cond}, top panel we show the differential conductance dI/dV versus V for a single lead set up analogous to Ref. \cite{Mourik_2012} where the current flows between a metallic lead to a semiconducting nanowire in proximity coupling to a superconducting lead. We have also calculated (not shown here) the same dI/dV versus V plots for a  set up involving two metallic leads separated by a topological superconducting nanowire. In both experiments, the signature of MZMs at the ends of the topological wire is a zero bias conductance peak which results from resonance local Andreev tunneling \cite{Law_Lee}. As shown in Fig. \ref{fig:cond} panel A, the same zero bias peak exists in the effectively topologically trivial regime as well with robust partially overlapping pairs of MBSs at each end resulting from long length scale potential inhomogeneity. Another \cite{Dassarma} signal which has been suggested recently as a signature of topologically protected Majorana modes, is the presence of correlated splitting oscillations as a function of an applied Zeeman field in measurements of LDOS.  In Fig. \ref{fig:LDOS} bottom panel we show that these zero energy oscillations appear in the trivial (yellow) region and persist into the topological (orange) regime without showing a signature due to the transition between the two regions.  We also confirm in Fig. \ref{fig:LDOS} bottom panel that these energy splittings are exponentially protected with increasing wire length in both the trivial and topological regimes.

In more recent work \cite{Marcus}, experiments on the semiconductor-superconductor heterostructures have been carried out in the Coulomb blockade limit, where the charging energy of the semiconductor segment discriminates between states with different numbers of electrons. In the set up with two metallic leads  and for $\Gamma > \Gamma_c$, ($\Gamma_c$ corresponds to Zeeman field for which the bulk excitation spectrum goes through a minimum, Fig. \ref{fig:FFig4}B, \ref{fig:FFig4}C) the zero bias conductance in the absence of charging energy is dominated by resonant Andreev tunneling through non-degenerate zero energy MZMs at each lead-superconductor interface \cite{Law_Lee}. In the Coulomb blockade limit with a finite charging energy $E_c$ ($E_c> \Delta > T$ where $\Delta$ is the proximity induced topological superconducting gap and $T$ is the temperature), the resonant Andreev tunneling is suppressed. A gate induced charge $N_g$ can be tuned, however, so that the ground state energy in the semiconductor island satisfies $E(N)=E(N+1)$, allowing the coherent transport of a single electron via the complex fermionic state $c,c^{\dagger}$ ($c=\gamma_1+i\gamma_2, c^{\dagger}=\gamma_1-i\gamma_2$) composed out of the MZMs localized near the ends. This process, also sometimes called teleportation \cite{Fu,Tewari_Zhang_Sarma,Sau_Swingle_Tewari}, is observable as a zero bias conductance peak periodic in the applied gate charge $N_g$ with a period $e$. In recent work \cite{Marcus}, observation of such a teleportation-like zero bias peak with increasing magnetic fields in InAs nanowire segments with epitaxial aluminum in the Coulomb blockade limit, as well as exponential protection of the zero energy states in the length of the segments, have been taken as strong evidence of the presence of MZMs.

In Fig. \ref{fig:cond} panel B we show the differential conductance as a function of bias potential and the Zeeman field for a two lead set up with the topological superconductor wire in the Coulomb blockade regime. In this regime, the charging energy $E_g$ is responsible for suppressing the anomalous tunneling (Andreev) processes at the lead-superconductor interfaces, as well as producing an effective overlap between the end localized MZMs even in the absence of a direct overlap of the wave functions \cite{Fu}. In the current BdG type calculations we incorporate the effect of the charging energy by manually suppressing the anomalous (Andreev) processes at each lead-superconductor interface. The remaining Majorana-assisted charge tunneling process between the two metallic leads on the two sides of the superconductor is expected to represent the teleportation amplitude. A check on our present calculations is the requirement that the zero bias peaks resulting from teleportation process should exponentially fall off with length of the wire because of a similar suppression of the wave function overlap. In Fig. \ref{fig:cond} panel C, we show the exponential fall-off of the zero bias conductance peaks in the Coulomb blockade regime for Zeeman splitting in the topological regime with two end localized  MZMs. We find a similar exponential fall off in the topologically trivial regime as well, with robust Andreev states due to potential fluctuations. In Fig. \ref{fig:cond} panel B , we show that pronounced zero bias peaks resulting from teleportation amplitude exists both in the topologically trivial (yellow) and non-trivial (orange) regimes in the presence of long length scale potential inhomogeneity.

In light of these results we conclude that demonstrating the nonlocal character of the topologically-protected MZMs and its emergence \textit{after} the system undergoes a quantum phase transition (or a crossover given by a minimum in the bulk excitation spectrum in finite systems), become critical tasks for the ongoing experimental search for MZMs in solid state heterostructures. In particular, we conclude that observing a zero-bias teleportation peak (at values of the gate charge with period $e$) that sticks to zero energy for a certain range of Zeeman fields, and exponential protection of the zero energy states on the length of the system revealed by splitting oscillations, \textit{do not} represent the unique signatures of topologically protected MZMs, because such signatures can also appear in the effectively topologically trivial phase via tunneling into partially overlapping pairs of MBSs near each end, which, despite being topologically trivial crosses over to behaving similar to robust MZMs for long length scale potential variations. Due to the fact that long length scale potential inhomogeneities arise generically in experimental systems with multiple gate-potentials, the ability to distinguish between partially overlapping MBSs within the topologically trivial regime and topologically protected MZMs in the topological regime becomes essential in the characterization of Majorana modes. A failure to take these potential inhomogeneities into account leaves claims of a unique or smoking gun signature \cite{Tripathi_Das,Valentini,Haim} of Majorana zero modes premature.

CM and ST thank ARO, and TDS was supported by NSF, DMR-1414683.



\end{document}